\documentclass[final,5p,times]{elsarticle}

\usepackage{lineno,hyperref}

\usepackage{amsmath}
\usepackage{amsfonts}
\usepackage{graphicx}
\usepackage{slashed}
 
\usepackage{feynmp-auto}
 
\newcommand{\dslash}{\partial\hskip-1.2ex/\hskip0.2ex}
\newcommand{\dmu}{\mathcal{D}}
\newcommand{\nn}{\nonumber}
\newcommand{\ordo}[1]{\mathcal{O}\left(#1\right)}
\newcommand{\Zsym}{$\mathbb{Z}_2$ }
\newcommand{\df}{\mathrm{d}}
 
\newcommand{\IE}{{\it i.e.~}}

\newcommand{\eq}[1]{eq.~\eqref{#1}} 

\renewcommand{\sec}[1]{section~\ref{#1}}

\newcommand{\mycites}[1]{refs.~\cite{#1}}
\newcommand{\mycite}[1]{ref.~\cite{#1}}

\usepackage{color}

\newcounter{bla}

\journal{Physics Letters B}

\begin{document}

\begin{frontmatter}

  \title{Scalar Kinetic Mixing and the Renormalization Group}
  
  \author{Johan Bijnens}\ead{bijnens@thep.lu.se}
  \author{Joel Oredsson}\ead{joel.oredsson@thep.lu.se}
  \author{Johan Rathsman}\ead{johan.rathsman@thep.lu.se}

  \address{Department of Astronomy and Theoretical Physics, Lund University,\\
   S\"olvegatan 14A, SE 223-62 Lund, Sweden}
  
  \begin{abstract}
    Quantum field theories containing scalar fields with equal quantum numbers allow for a mixed kinetic term in the Lagrangian. 
    It has been argued that this mixing must be taken into consideration when performing renormalization group (RG) analyses of such a theory.
    However, from the fact that scalar kinetic mixing does not correspond to a physical observable, we show that no extra parameters need to be introduced.
    Using a toy model, we explicitly derive the 1-loop RG equations (RGEs) in three different renormalization schemes to demonstrate how this issue can be dealt with.
    In schemes without kinetic mixing, either the fields mix during renormalization to produce non-diagonal anomalous dimensions or the RGEs explicitly depend on the scalar masses.
    Finally, we show how the different schemes are related to each other by scale dependent field redefinitions. 
  \end{abstract}
  
  
  \end{frontmatter}

\section{Introduction}

One common feature of beyond the standard models is the inclusion of extra scalar fields. 
A very well studied example is the so called two Higgs Doublet Model (2HDM) which was introduced in 1974 by T.\ D.\ Lee \cite{Lee:1973iz}. 
As its name implies, it has two Higgs doublets instead of one as in the standard model. 
Most phenomenological studies of the 2HDM assumes that the two Higgs doublets are charged differently under a discrete \Zsym symmetry and that this symmetry is at most softly broken, \IE by mass-terms, in the scalar potential. 
By assigning appropriate charges to the right-handed fermion fields, this ensures that there are no tree-level Flavor Changing Neutral Currents (FCNC). 
However, in general there is no justification to make this assumption, since the 2HDM in itself is also not a complete theory. 
For example having the same fine-tuning or hierarchy problem as the standard model. 
In fact, in a supersymmetric version of the theory, the \Zsym symmetry is broken when going beyond tree-level \cite{Hall:1985dx} and even though this breaking is only soft it gives rise to so called non-holomorphic corrections. 

In this paper we want to clarify in a pedagogical way what happens when the \Zsym symmetry is broken. 
One immediate consequence is that the two Higgs doublets will mix, giving rise to both kinetic and mass mixing. 
At tree level, as is well known\footnote{See for example sec.12.5 in \mycite{Weinberg:1995mt}.}, the kinetic term as well as the mass matrix can be diagonalized. 
The question we investigate in this paper is whether this can be done consistently also at loop-level, \IE after renormalization, and whether the kinetic term can be kept diagonal also under renormalization group evolution, even though the fields mix. 
In \mycites{Ginzburg:2004vp,Ginzburg:2008kr} it was claimed that whereas one can always make the kinetic term diagonal at a given renormalization scale, the mixing will reappear if the renormalization scale is changed, thus meaning that one gets an additional parameter. 
This claim has also led to some confusion in the literature \cite{Gorda:2018hvi}.
As we will show in this paper, this claim is not correct and one does not need an additional parameter in order to renormalize the theory. 
Instead, one can choose to work with different renormalization schemes where this parameter is either present or not, all giving the same results. 
The underlying reason behind this result is that the two-point functions are in fact not observables and as such do not need to be finite; the observables are masses and scattering amplitudes. 
We also show that in those schemes where there is no kinetic mixing, one instead has mixing of the fields under renormalization corresponding to a change of basis which has to be taken into account.

A similar problem occurs in the renormalization of the CKM matrix.
For a recent discussion of how that is resolved in different renormalization schemes for the case of an extended scalar sector, see \mycite{Denner:2018opp}.

We present three different renormalization schemes with or without kinetic mixing and show how they are related at one-loop level under renormalization through orthogonal and non-orthogonal transformations. 
A difference compared to tree-level is that these transformations are scale-dependent. 
As a by-product of our considerations we also show that in one of these schemes, the renormalization group equations in the Minimal Subtraction (MS) scheme for the quartic couplings depend on the masses of the scalar particles - in contradiction with statements often made in textbooks. 

As a pedagogical example and to simplify the discussions, we will not consider the full 2HDM but instead only consider a toy model with $N \ge 2$ real pseudo scalar fields coupled to one Dirac fermion. 
This model has the same generic features as the 2HDM when it comes to renormalization properties of the scalar fields.

This paper is organized as follows: We start in section 2 by defining the theory and introduce the three different renormalization schemes that we are considering. 
In the following section we go through each scheme in more detail, perform the renormalization at one-loop level and calculate the RGEs for each case. 
In section 4 we then show how the schemes are related to each other and that they in fact are equivalent. Finally section 5 contains our conclusions.

\section{Kinetic mixing and renormalizability}
\label{kineticmixing}

To discuss the renormalizability of theories with mixed scalar kinetic terms, we have chosen the simplest case with $N$ real pseudo scalar fields coupled to a Dirac fermion.
The reason to include a fermion is because then the 2-point Green functions,
\begin{align}
	G_{ij}(p^2)=\vcenter{\hbox{
	\begin{fmffile}{Gij}
		\begin{fmfgraph*}(40,20)
			\fmfleft{i1}
			\fmfright{o1}
			\fmf{dashes}{i1,v1}
			\fmfblob{0.4w}{v1}
			\fmf{dashes}{v1,o1}
			\fmflabel{$i$}{i1}
			\fmflabel{$j$}{o1}
		\end{fmfgraph*}
	\end{fmffile}
	}}
	~=\frac{1}{\epsilon}\left(\alpha_{ij} p^2 + \beta_{ij}\right),
\end{align}
exhibit divergences proportional to $p^2$ at 1-loop level; instead of at 2-loop level, which would be the case in a theory with only scalars.

The only requirement for a theory to be renormalizable is that all physical quantities are free from divergences. 
So even if some component of $G_{ij}(p^2)$ contains divergences, it in itself does not spoil renormalizability since $G_{ij}(p^2)$ is not a physical observable.
In our theory, the physical observables are scattering amplitudes and masses of the particles.
Thus, the minimal set of parameters and counterterms consists only of interaction and diagonal mass terms.
We will show in \sec{sectminimal} how this set is sufficient to absorb all divergences.
This merely corresponds to one particular renormalization scheme.

Another renormalization scheme is considered in \sec{sectgeneral}, where we work with the most general Lagrangian consistent with the symmetries of the theory, including kinetic mixing operators, and renormalize every parameter.
Since the parameters in the Lagrangian are not directly related to any physical observables, one instead requires all Green functions to be free of divergences.

A third renormalization scheme, and the one most often used, is to work with diagonal kinetic terms and renormalized fields.
To be able to absorb the divergences in all Green functions into counterterms, the scalar fields must mix during renormalization.
This will induce the necessary non-diagonal kinetic counterterms to ensure renormalizability.
The anomalous dimensions of the fields will then enter the RGEs for the couplings.
We show how this is done in \sec{sectstandard}.

Throughout all of our calculations, we use dimensional regularization in $d=4-2\epsilon$ dimensions and MS for the counterterms.
We will also introduce the renormalization scale $\mu$ to make all the couplings in the Lagrangian have their natural dimension.
All bare quantities that are $\mu$ independent will be denoted by a $b$ superscript.
For notational convenience, we define $\Lambda \equiv 1/(16\pi^2 \epsilon)$ and $\dmu \equiv 16\pi^2 \mu\frac{\df}{\df\mu}$ as well as
\begin{align}\label{GreenFuncs}
  i\Sigma(\slashed{p})=~&
  \vcenter{\hbox{
  \begin{fmffile}{Sigma}
    \begin{fmfgraph*}(40,20)
      \fmfleft{i1}
      \fmfright{o1}
      \fmf{fermion}{i1,v1}
      \fmfpolyn{empty,smooth,label=1PI,tension=0.4}{v}{4}
      \fmf{fermion}{v3,o1}
    \end{fmfgraph*}
  \end{fmffile}
  }}, &
  i\Pi_{ij}(p^2)=~&\vcenter{\hbox{
  \begin{fmffile}{Piij}
    \begin{fmfgraph*}(40,20)
      \fmfleft{i1}
      \fmfright{o1}
      \fmf{dashes}{i1,v1}
      \fmfpolyn{empty,smooth,label=1PI,tension=0.4}{v}{4}
      \fmf{dashes}{v3,o1}
      \fmflabel{$i$}{i1}
      \fmflabel{$j$}{o1}
    \end{fmfgraph*}
  \end{fmffile}
  }}\nn\\\nn\\
  \Gamma_i^{\text{amp}}=~&\left(\vcenter{\hbox{
  \begin{fmffile}{Gi}
    \begin{fmfgraph*}(40,30)
      \fmfleft{i1,i2}
      \fmfright{o1}
      \fmf{fermion}{i1,v1,i2}
      \fmfblob{0.4w}{v1}
      \fmf{dashes}{v1,o1}
      \fmflabel{$i$}{o1}
    \end{fmfgraph*}
  \end{fmffile}
  }}
  ~\right)_{\text{amp}}, &
  \Gamma_{ijkl}^{\text{amp}}=~&\left(\vcenter{\hbox{
  \begin{fmffile}{Gijkl}
    \begin{fmfgraph*}(40,30)
      \fmfleft{i1,i2}
      \fmfright{o1,o2}
      \fmf{dashes}{i1,v1,o1}
      \fmf{dashes}{i2,v1,o2}
      \fmfblob{0.4w}{v1}
      \fmflabel{$i$}{i2}
      \fmflabel{$j$}{i1}
      \fmflabel{$k$}{o2}
      \fmflabel{$l$}{o1}
    \end{fmfgraph*}
  \end{fmffile}
  }}
  \right)_{\text{amp}} .
\end{align}

\section{Three renormalization schemes}

\subsection{Most general case}
\label{sectgeneral}

The most general Lagrangian with the imposed symmetries in terms of bare fields and parameters is given by
\begin{align}
\label{Lgeneral}
\mathcal{L} =& \frac{1}{2}\partial_\mu\varphi_i^b a_{ij}^b
               \partial^\mu \varphi_j^b
- \frac{1}{2}\varphi_i^bb_{ij}^b \varphi_j^b
+ c^b \overline\psi^b i\dslash\psi^b
- d^b\overline\psi^b\psi^b
\nonumber\\
&- e_i^b \varphi_i^b \overline\psi^b i\gamma_5\psi^b
-\frac{1}{4!} f_{ijkl}^b\varphi_i^b\varphi_j^b\varphi_k^b\varphi_l^b\,.
\end{align}
The bare parameters are divided into renormalized parameters and counter terms as
\begin{align}
\label{baregeneral}
 a_{ij}^b =& a_{ij}+\delta a_{ij} ,
&
 b_{ij}^b =& b_{ij}+\delta b_{ij},
&
 c^b =& c+\delta c,
\nonumber\\ 
d^b =& d+\delta d,
&
e_i^b =& \mu^\epsilon\left(e_i+\delta e_i\right),
&
f_{ijkl}^b =& \mu^{2\epsilon}\left(f_{ijkl}+\delta f_{ijkl}\right),
\end{align}
where $a_{ij},b_{ij},f_{ijkl}$ are fully symmetric in all their indices and appropriate powers of $\mu$ have been inserted to give all terms correct dimensions.
We will also use matrix and vector notation for the coefficients
$a_{ij},b_{ij},e_i$.  

This Lagrangian contains a number of  redundant parameters. 
In the case of two scalar fields, we have 15 free parameters with their associated counterterms.
These do not directly correspond to physical observables.  
If one transforms the Lagrangian to the minimal form, as we do in \sec{sectminimal}, the total number of free parameters is reduced to 10; which then directly correspond to physical observables.

One complication of working with this general Lagrangian is the non-diagonal kinetic terms; which gives rise to a matrix scalar propagator.
In the calculations we treat the mass parameters as small perturbations  $b_{ij} \ll p^2 $ and work with an expanded propagator,
\begin{align}\label{MatrixPropagator}
	\vcenter{\hbox{
	\begin{fmffile}{ScalarProp}
		\begin{fmfgraph*}(40,20)
			\fmfleft{i1}
			\fmfright{o1}
			\fmf{dashes}{i1,o1}
			\fmflabel{$i$}{i1}
			\fmflabel{$j$}{o1}
		\end{fmfgraph*}
	\end{fmffile}
	}}
	~=\left(\frac{i}{p^2 a-b}\right)_{ij}\simeq \frac{i}{p^2}\left(a^{-1}+\frac{a^{-1}ba^{-1}}{p^2}\right)_{ij}.
\end{align}

Ignoring the finite pieces, the 1-loop calculation of correlation functions give
\begin{align}
	\Sigma(\slashed{p}) =& \left( \frac{e_ie_ja_{ij}^{-1}}{2c}\slashed{p}-\frac{e_ie_ja_{ij}^{-1}}{c^2}d\right)\Lambda+(\slashed{p}\delta c-\delta d),\\
  \Pi_{ij}(p^2) =& \left[\frac{1}{2}f_{ijkl}(a^{-1}ba^{-1})_{kl}+(4d^2-2c^2p^2)\frac{e_ie_j}{c^4}\right]\Lambda \nn\\
    &+ (p^2\delta a_{ij}-\delta b_{ij}),
\end{align}
\begin{align}
	\Gamma_i^{\text{amp}} =& -\frac{e_ie_je_ka_{jk}^{-1}}{c^2}\Lambda +\delta e_i,\\
	\Gamma_{ijkl}^{\text{amp}} =& \left[\frac{1}{2}\left(a^{-1}\right)_{mn}
        \left(a^{-1}\right)_{op}\left(f_{ijmo}f_{klnp}+f_{ikmo}f_{jlnp}+f_{ilmo}f_{jknp}\right)\right.\nn\\
        &\left.-\frac{24 e_ie_je_ke_l}{c^4}
\right]i\Lambda - i\delta f_{ijkl}.
\end{align}
As renormalization conditions we will require all Green functions to be be finite; which is accomplished by absorbing all the infinites in the 1PI and amputated diagrams into the counterterms. 
Requiring the bare fields and parameters in \eq{baregeneral} to be $\mu$-independent leads straightforwardly to the RGEs for the renormalized ones.
In four dimensions they are
\begin{align}
\label{RGEgeneral}
\dmu a_{ij} =& \frac{4e_ie_j}{c^2},
\nonumber\\
\dmu b_{ij} =& \frac{8d^2e_i e_j}{c^4}+f_{ijkl}\left(a^{-1}ba^{-1}\right)_{kl},
\nonumber\\
\dmu c =& -\frac{\left(e^T a^{-1} e\right)}{c},
\nonumber\\
\dmu d =& -\frac{2\left(e^T a^{-1}e\right)d}{c^2},
\nonumber\\
\dmu e_i =& \frac{2e_i}{c^2}\left(e^T a^{-1} e\right),
\nonumber\\
\dmu f_{ijkl} =& \left(a^{-1}\right)_{mn}
        \left(a^{-1}\right)_{op}\left(f_{ijmo}f_{klnp}+f_{ikmo}f_{jlnp}+f_{ilmo}f_{jknp}\right)\nn\\
        &-\frac{48 e_ie_je_ke_l}{c^4}.
\end{align}

From the first equation above we see that in this renormalization scheme there is indeed kinetic mixing if the fermion couples to more than one of the scalar fields; much like the scheme used in \mycite{Ginzburg:2008kr}.
However, not all of the parameters above are physical and as we will see later the parameter describing kinetic mixing is redundant.

\subsection{Minimal case}
\label{sectminimal}

Here we only add the minimal number of free parameters and counterterms.
This is similar to what is used in effective field theory methods where
one removes the so-called equations of motion terms or alternatively
brings the Lagrangian into the minimal form using field redefinitions
\footnote{See, for example, sec.\ 6 in \mycite{Manohar:2018aog} for a pedagogical introduction.}.
The physical observables in our theory are the masses and scattering amplitudes; therefore we only require counterterms for the masses and couplings.
The Lagrangian in terms of the bare quantities is
\begin{align}
\label{Lminimal}
\mathcal{L} =&
\frac{1}{2}\partial_\mu\Phi_i^b\partial^\mu\Phi_i^b
-\frac{1}{2} m_i^{2b} \Phi_i^b\Phi_i^b
+\overline\Psi^b i\dslash\Psi^b
-m^b\overline\Psi^b\Psi^b
\nonumber\\&
-Y_i^b \Phi_i^b \overline\Psi^b i\gamma_5\Psi^b
-\frac{1}{4!}\Lambda_{ijkl}^b
  \Phi_i^b\Phi_j^b\Phi_k^b\Phi_l^b\, ,
\end{align}
where $\Lambda_{ijkl}^b$ is fully symmetric in $i,j,k,l$.
The bare parameters are related to the renormalized ones and counterterms via
\begin{align}
\label{bareminimal}
m_i^{2b} =& m_i^2+\delta m_i^2,
&
m^b =& m+\delta m,
\nonumber\\
Y_i^b =& \mu^\epsilon \left(Y_i+\delta Y_i\right),
&
\Lambda_{ijkl}^b =& \mu^{2\epsilon}\left(\Lambda_{ijkl}+\delta\Lambda_{ijkl}\right).
\end{align}
Here we only have the physically relevant number of parameters and the
same number of counterterms. 
For the case of two scalars this would be 10, although we will work with any number of scalars.
We will also assume the masses to be non-
degenerate in order to keep the presentation simple.
In the case of degenerate masses, the conclusions
below do not change but the analysis needs to use methods
from degenerate perturbation theory to avoid singularities in
e.g.\ \eq{defCij}.

In the minimal scheme, the UV divergent pieces and counterterms of the 1PI and amputated Green functions in \eq{GreenFuncs} are
\begin{align}
	\Sigma(p^2) =& \sum_iY_i^2\left(\frac{\slashed{p}}{2}-m\right)\Lambda-\delta m,\\
	\Pi_{ij}(p^2) =& \left[\frac{1}{2}\Lambda_{ijkk}m_k^2+\frac{1}{2}Y_iY_j(8m^2-4p^2)\right]\Lambda-\delta_{ij}\delta m_i^2,\\	
	\Gamma_i^{\text{amp}} =& -Y_i \sum_j Y_j^2 \Lambda + \delta Y_i,\\
	\Gamma_{ijkl}^{\text{amp}} =& \left[\frac{1}{2}
  \left(\Lambda_{ijmm}\Lambda_{klmm}+\Lambda_{ikmm}\Lambda_{jlmm}+\Lambda_{ilmm}\Lambda_{jkmm}\right)\right.\nn\\
  &\left.-24Y_iY_jY_kY_l\right]i\Lambda-i\delta\Lambda_{ijkl}.
\end{align}
Summing all 1PI diagrams gives the full 2-point Green function of bare fields,
\begin{align}
  G_{ij}(p^2)\equiv & \left(\frac{i}{p^2-M+\Pi(p^2)}\right)_{ij}, 
\end{align}
where $M_{ij}=m_i^2\delta_{ij}$.
To 1-loop order it reduces to
\begin{align}
	G_{ij}=\left\{\begin{array}{lr} \dfrac{i}{p^2-m_i^2}\left(1-\dfrac{\Pi_{ii}}{p^2-m_i^2}\right), & \text{for }i=j,\\ \\
									\dfrac{-i\Pi_{ij}}{(p^2-m_i^2)(p^2-m_j^2)}, & \text{for }i\neq j.\end{array}\right.
\end{align}

A crucial observation is that one cannot make all components of $G_{ij}$ to be finite since we cannot absorb the off-diagonal $p^2$ divergent piece in $\Pi_{ij}$ in any of the counterterms.
However, this is not a problem since 2-point Green functions are not observables in quantum field theory and can thus contain divergences.
The relevant physical observables are the poles of the 2-point functions, \IE the masses of the particles,
as well as the S-matrix elements, related to correlation functions through the LSZ theorem \cite{1955NCimS...1..205L}.
The masses are fixed by $\det G^{-1}=0$ on-shell and to make them finite, the $\delta m_i^2$ will absorb all the infinities in the corresponding eigenvalue; which to 1-loop order are the ones in $\Pi_{ii}(p^2=m_i^2)$.

In a theory where external particles have the same quantum numbers, the conventional LSZ theorem needs to be modified to include mixing on the legs \cite{Amoros:2001cp}.
The S-matrix element describing scattering of scalar mass eigenstates $ij$ going into $kl$ is then given by\footnote{The masses in the LSZ theorem are the physical pole masses, which are equivalent to the MS masses up to finite pieces. Since we are only discussing UV divergences, we will not make any distinction between pole masses and MS masses. The difference would need to be taken into consideration at higher orders in perturbation theory.} 
\begin{align}
	\mathcal{A}_{ijkl} = &\lim_{p_1^2\rightarrow m_i^2} (p_1^2-m_i^2)G_{ia}(p_1^2)
	\times\lim_{p_2^2\rightarrow m_j^2} (p_2^2-m_j^2)G_{jb}(p_2^2)\nn\\
	&\times\lim_{p_3^2\rightarrow m_k^2} (p_3^2-m_k^2)G_{kc}(p_3^2)
	\times\lim_{p_4^2\rightarrow m_l^2} (p_4^2-m_l^2)G_{l d}(p_4^2)\nn\\
	&\times \frac{1}{\sqrt{Z_{ii}^{(i)}Z_{jj}^{(j)}Z_{kk}^{(k)}Z_{ll}^{(l)}}} \Gamma^{\text{amp}}_{abcd}(p_1,p_2,p_3,p_4),
\end{align}
where the $Z_{ij}^{(k)}$ factors are defined as the residues of the corresponding 2-point correlation functions,
\begin{align}
	G_{ij}(p^2) \rightarrow \frac{i Z_{ij}^{(k)}}{p^2-m_k^2+i\epsilon}+\ordo{1},
\end{align}
as $p^2\rightarrow m_k^2$, \IE
\begin{align}
	Z_{ij}^{(k)} =-i\lim_{p^2\rightarrow m_k^2}(p^2-m_k^2)G_{ij}(p^2).
\end{align}
Taking the external momenta on-shell, the amplitude becomes
\begin{align}
	\mathcal{A}_{ijkl} =&\left(1-\frac{1}{2}\sum_{a=i,j,k,l}\frac{d}{dp^2}\Pi_{aa}\Big|_{p^2=m_a^2}\right)\Gamma_{ijkl}^{\text{amp}}\nn\\
		&+\sum_{a\neq i} \frac{\Pi_{ia}(p^2=m_i^2)}{m_a^2-m_i^2}\Gamma_{ajkl}^{\text{amp}}+\sum_{a\neq j} \frac{\Pi_{ja}(p^2=m_j^2)}{m_a^2-m_j^2}\Gamma_{iakl}^{\text{amp}}\nn\\
		&+\sum_{a\neq k} \frac{\Pi_{ka}(p^2=m_k^2)}{m_a^2-m_k^2}\Gamma_{ijal}^{\text{amp}}+\sum_{a\neq l} \frac{\Pi_{la}(p^2=m_l^2)}{m_a^2-m_l^2}\Gamma_{ijka}^{\text{amp}}.
\end{align}
A similar discussion applies when considering $\Phi_i-\bar{\Psi}\Psi$ scattering and here we only write the final expression in terms of 1PI and amputated diagrams,
\begin{align}
  \mathcal{A}_i =& \left(1-\frac{d}{d\slashed{p}} \Sigma(\slashed{p})\Big|_{\slashed{p}=m}-\frac{1}{2}\frac{d}{dp^2}\Pi_{ii}\Big|_{p^2=m_i^2}\right)\Gamma_i^{\text{amp}}\nn\\
  &+\sum_{a\neq i}\frac{\Pi_{ia}(p^2=m_i^2)}{m_a^2-m_i^2}\Gamma_a^{\text{amp}}.
\end{align}

As renormalization conditions we now require all observables to be finite, \IE $\mathcal{A}_i$, $\mathcal{A}_{ijkl}$ and the masses $m_i^2$, $m$.
The infinities in $G_{ij}$ that can not be absorbed into counterterms from $\Pi_{ij}$ will then be canceled in the observables by the counterterms $\delta Y_i$ and $\delta\Lambda_{ijkl}$.

Note that other observables like $\Phi_i\Psi\to\Phi_j\Psi$ and $\Psi\Psi\to\Psi\Psi$ will also be finite. 
Even though they contain off-shell subdiagrams that contain divergent pieces; these will cancel in the sum of all the amplitudes contributing to a certain process.

Since all the fields are bare and do not depend on the renormalization scale, the RGEs for the couplings and masses follow straightforwardly from the counterterms\footnote{There is no implicit sum over repeated indices in \eq{RGEminimal}.}:
\begin{align}
\label{RGEminimal}
\dmu m_i^2 =& \sum_j\Lambda_{iijj}m_j^2+8Y_i^2 m^2 -4 Y_i^2 m_i^2,
\nonumber\\
\dmu m =& -m \sum_iY_i^2,
\nonumber\\
\dmu Y_i =& -2Y_i^3+3Y_i\sum_kY_k^2+2\sum_{l\ne i}Y_l C_{il},
\nonumber\\
\dmu\Lambda_{ijkl}=&
-2\left(Y_i^2+Y_j^2+Y_k^2+Y_l^2\right)\Lambda_{ijkl}-48Y_iY_jY_kY_l\nn\\
&+\sum_m(\Lambda_{ijmm}\Lambda_{klmm}+\Lambda_{ikmm}\Lambda_{jlmm}+\Lambda_{ilmm}\Lambda_{jkmm})
\nn\\
&+2\sum_{m\ne i}\Lambda_{mjkl}C_{im}
+2\sum_{m\ne j}\Lambda_{imkl}C_{jm}\nn\\
&+2\sum_{m\ne k}\Lambda_{ijml}C_{km}
+2\sum_{m\ne l}\Lambda_{ijkm}C_{lm},
\end{align}
where we have defined
\begin{align}\label{defCij}
C_{il}\equiv&\frac{1}{m_i^2-m_l^2}\left(\sum_k\frac{\Lambda_{ilkk}}{2}m_k^2+4Y_iY_l m^2-2 Y_i Y_l m_i^2\right).
\end{align} 
It should be noted again that we assume the scalar masses to be non-degenerate. 
It is interesting and unusual that the RGEs in this MS renormalization scheme depend on the masses through the $C_{ij}$ terms.
However, one could expect this; since this scheme is related to an on-shell renormalization one.
It should also be noted that the definition of the fields in this case is unchanged during the RG evolution; they are always given by the mass eigenstates. 
In other words, the basis is the same throughout the evolution. 
Another point is that the contributions from the off-diagonal $C_{ij}$-terms vanish for two scalars if a \Zsym symmetry is imposed. 
In other words these terms contain the same information as the kinetic mixing terms in \sec{sectgeneral}.

\subsection{Standard way with $Z_{ij}$}
\label{sectstandard}

The standard renormalization scheme most often used is very similar to the case of working with a completely general Lagrangian as in \sec{sectgeneral}.
But one obvious simplification is to transform the fields to arrive at canonical kinetic terms through a non-orthogonal\footnote{In more general cases with complex fields it would require a non-unitary transformation.} transformation.
For this scheme we will also renormalize the fields, which gives rise to anomalous dimensions that will enter the RGEs for all the parameters.

We denote bare fields by $\phi_i^b$, $\overline{\Psi^b}$, $\Psi^b$ and renormalized fields by $\phi_i, \overline\Psi,\Psi$.
The relations between the two are given by $\phi_i^b = Z_{ij}\phi_j$ and $\Psi^b = Z_\Psi \Psi$.
With the renormalization factors $Z_{ij} = \delta_{ij}+\delta Z_{ij}$ and $Z_\Psi = 1+\delta\Psi$.
The Lagrangian in terms of renormalized quantities and counterterms is then given by
\begin{align}
\label{Lstandard}
\mathcal{L}=&
\frac{1}{2}Z_{ik}\partial_\mu\phi_k Z_{il}\partial^\mu\phi_l
-\frac{1}{2}Z_{ik}\phi_k\left( m_{ij}^2+\delta m_{ij}^2\right) Z_{jl}\phi_l\nn\\
&+Z_\Psi^2\overline\Psi i\dslash\Psi-Z_\Psi^2 \left(m+\delta m\right)\overline\Psi\Psi
-\mu^\epsilon\left(y_i+\delta y_i\right)\phi_i\overline\Psi i\gamma_5\Psi \nn\\
&-\frac{\mu^{2\epsilon}}{4!}\left(\lambda_{ijkl}+\delta\lambda_{ijkl}\right)\phi_i\phi_k\phi_k\phi_l,
\end{align}
where $m_{ij}^2$ and $\lambda_{ijkl}$ are fully symmetric in their indices. 
Note that we have a different number of parameters compared to counterterms here.
For two scalars we have 15 counterterms and 11 parameters, but of course only 10 physical parameters as in the other schemes.
The relations of bare and renormalized parameters are somewhat more complicated than in the previous renormalization schemes, 
\begin{align}
{m_{ij}^b}^2 =~ m_{ij}^2+\delta m_{ij}^2 , &&&
y_i^b =~ Z_{ji}^{-1}Z_\Psi^{-2}\mu^\epsilon(y_j+\delta y_j),
\end{align}
\begin{align}
\lambda_{ijkl}^b =~& Z_{ai}^{-1}Z_{bj}^{-1}Z_{ck}^{-1}Z_{dl}^{-1}\mu^{2\epsilon}(\lambda_{abcd}+\delta\lambda_{abcd}). 
\end{align}
Since the renormalization factors are $\mu$ dependent we now obtain RGEs for the fields,
\begin{align}
	\dmu \phi_i =& -\left( Z^{-1} \dmu Z\right) _{ij} \phi_j \equiv \gamma_{ij} \phi_j,\nn\\
	\dmu \Psi =& - Z_\Psi^{-1} \dmu Z_\Psi \Psi \equiv \gamma_\Psi \Psi.
\end{align}
In case of $\gamma_{ij}$ not being diagonal this means that the fields will mix during renormalization and thereby the basis will also change. 
These anomalous dimensions, $\gamma_{ij}$ and $\gamma_\Psi$, also enter the RGEs for the parameters
\begin{align}
	\dmu y_i =& -(4\pi)^2\epsilon(y_i+\delta y_i) -\gamma_{ij}y_j-2\gamma_\Psi g_i -\dmu \delta y_i,\\
	\dmu \lambda_{ijkl} =& -2(4\pi)^2\epsilon(\lambda_{ijkl}+\delta\lambda_{ijkl})-\dmu \delta\lambda_{ijkl}\nn\\
	&-\sum_m (\gamma_{im}\lambda_{mjkl} +\gamma_{jm}\lambda_{imkl}+\gamma_{km}\lambda_{ijml} +\gamma_{lm}\lambda_{ijkm}).
\end{align}

Just as in \sec{sectgeneral}, the counterterms are determined by requiring that all Green functions of the
fields $\phi_i,\overline\Psi,\Psi$ are finite.
This is accomplished by making the 1PI and amputated diagrams in \eq{GreenFuncs} finite, which we compute to be
\begin{align}
	\Sigma(p^2) =& \sum_iy_i^2\left(\frac{\slashed{p}}{2}-m\right)\Lambda+(2\slashed{p}\delta\Psi-2m\delta\Psi-\delta m),\\
  \Pi_{ij}(p^2) =& \left[\frac{1}{2}\lambda_{ijkl}m_{kl}^2+y_iy_j(4m^2-2p^2)\right]\Lambda\nn\\
  &+(2p^2\delta Z_{ij}-m_{ja}^2\delta Z_{ia}-m_{ia}^2\delta Z_{ja}-\delta m_{ij}^2),
\end{align}
\begin{align}
	\Gamma_i^{\text{amp}} =& -y_i \sum_j y_j^2 \Lambda + \delta y_i,\\
	\Gamma_{ijkl}^{\text{amp}} =& \left[\frac{1}{2}
  \left(\lambda_{ijmm}\lambda_{klmm}+\lambda_{ikmm}\lambda_{jlmm}+\lambda_{ilmm}\lambda_{jkmm}\right)\right.\nn\\
  &\left.-24y_iy_jy_ky_l\right]i\Lambda-i\delta\lambda_{ijkl}.
\end{align}
In the end, we arrive at the following anomalous dimensions and RGEs for the parameters
\begin{align}\label{RGEstandard}
	\gamma_{ij} =& 2y_iy_j,\nn\\
	\gamma_\Psi =& -\frac{1}{2}\sum_i y_i^2,\nn\\
\dmu m_{ij}^2=&
\lambda_{ijkl}m_{kl}^2+8m^2 y_iy_j-2y_iy_km_{kj}^2-2m_{ik}^2y_ky_j,
\nonumber\\
\dmu m=&
-  m \sum_i y_i^2,
\nonumber\\
\dmu y_i=& y_i \sum_k y_k^2,
\nonumber\\
\dmu \lambda_{ijkl} =&
\lambda_{ijmn}\lambda_{klmn}+\lambda_{ikmn}\lambda_{jlmn}
 +\lambda_{ikmn}\lambda_{jlmn}-48 y_i y_j y_k y_l
\nonumber\\&
-2y_iy_m\lambda_{mjkl}
-2y_jy_m\lambda_{imkl}-2y_ky_m\lambda_{ijml}-2y_ly_m\lambda_{ijkm}.
\end{align}
Here, we again see that if one fermion couples to more than one of the scalars, then the scalars will mix under RG evolution from $y_iy_j \neq 0$ for $i\neq j$. The effect appears both in the anomalous dimensions of the scalar fields as well as in the masses and quartic couplings. This then corresponds to the kinetic mixing in \sec{sectgeneral}. 
An example of how to actually perform the RG evolution with these effects taken into consideration can be found in \mycite{Oredsson:2018yho}.

As a side note we briefly compare these results to the renormalization scheme used in \mycites{Machacek:1983tz,Machacek:1983fi,Machacek:1984zw,Luo:2002ti}, where they derive the 1- and 2-loop RGEs for a general quantum field theory.
There is a subtlety in their notation in that they present the RGEs for a theory with an irreducible representation of the scalar fields and the anomalous dimensions are therefore taken to be diagonal
\footnote{This has independently been pointed out by \mycite{Bednyakov:2018cmx}.}.
Their formulas can however be modified to include theories containing multiple scalar fields by generalizing the anomalous dimensions in a relatively straightforward way.
This is discussed in great detail in \mycite{Schienbein:2018fsw} and we will not discuss it further here.

\section{Relation between the various schemes}

All the renormalization schemes are built on bare Lagrangians, which of course do not depend on the renormalization scale $\mu$.
The different bare Lagrangians are then related to each other by field redefinitions and therefore the renormalized fields are as well.
In this section we will show that the renormalized parameters in each of the three renormalization schemes are related to each other by orthogonal and non-orthogonal transformations. 
Though the transformations are somewhat trivial at bare level, the transformation matrices of renormalized quantities do obey a non-trivial $\mu$-dependence.

\subsection{Most general versus minimal}
\label{GeneralvsMinimal}

To relate the general Lagrangian in \sec{sectgeneral} to the minimal Lagrangian in \sec{sectminimal} we need to first diagonalize and normalize the kinetic terms, \IE $a_{ij}$.
Following that, we need another rotation to diagonalize the new mass matrix.
We write the full non-orthogonal transformation matrix as
\begin{align}
\label{eq:defRb}
R^b =& O_1^{bT} {A^b}^{-1} O_2^{bT},
\end{align}
where the diagonal matrix $A^b$ and orthogonal matrices $O_i^b$ are defined by
\begin{align}\label{eq:defRb2}
O_1^b a^b O_1^{bT} = &\mathrm{diag}(\alpha_i^b),
&
A^b =&\mathrm{diag}\left(\sqrt{\alpha_i^b}\right),
\nn\\
O_2^b{A^b}^{-1}O_1^b b^b O_1^{bT}{A^b}^{-1}O_2^{bT} =&
\mathrm{diag}\left(m_i^{2b}\right),
\end{align}
such that
\begin{align}\label{eq:defRb3}
R^{bT} a R^b =& 1,
& 
R^{bT} b^b R^b =
\mathrm{diag}\left(m_i^{2b}\right) \equiv M^b.
\end{align}
The relations of the bare fields are then
\begin{align}
\label{eq:relfields}
\varphi_i^b = R_{ij}^b \Phi_j^b
&&\text{and}&&
\psi^b = \frac{1}{\sqrt{c^b}}\Psi^b.
\end{align}
Note that since the transformation matrix $R^b$ contains divergences, it is clear that Green functions of
$\Phi_i^b,\overline\Psi^b,\Psi^b$ can be divergent while those
of $\phi_i^b,\overline\psi^b,\psi^b$ are finite.

While the relation between $m_i^{2b}$ and $a^b,b^b$ is implicit in eqs.\ (\ref{eq:defRb}--\ref{eq:defRb3}),
the other bare parameters are related via
\begin{align}
\label{eq:relbare}
m^b =& \frac{d^b}{c^b},
&
Y_i^b =& \frac{1}{c^b}e_j^b R_{ji}^b,
&
\Lambda_{ijkl}^b =& f_{abcd}^b R_{ai}^b R_{bj}^b R_{ck}^b R_{dl}^b.
\end{align}

All quantities involved in eqs.\ (\ref{eq:defRb}--\ref{eq:relbare}) are bare quantities
and are $\mu$-independent; as a consequence, the counterterms and RGEs must also be compatible.
We can then relate the renormalized quantities with a $\mu$-dependent transformation matrix $R_{ij}(\mu)$.
The relations between the parameters in the two schemes are the same as in the bare case, but without the superscript $b$ everywhere.
Using the notation $M=\mathrm{diag}(m_i^2)$ these are
\begin{align}
\label{defR}
R^T a R =& 1,
&
R^T b R =& M,
&
\frac{d}{c} =& m,
\nonumber\\
\frac{1}{c} e^T R=&Y^T,
&
f_{abcd}R_{ai} R_{bj} R_{bk} R_{dl} =& \Lambda_{ijkl}.
\end{align}
To figure out how $R_{ij}$ depends on the renormalization scale, one can determine its explicit form in terms of the renormalized Lagrangian parameters; which of course fully fixes the $\mu$-dependence.
Another way is using the relations in \eq{defR} to transform the RGEs in the general scheme in \eq{RGEgeneral} to the RGEs in the minimal scheme in \eq{RGEminimal} from which the required $\mu$-dependence for $R_{ij}$ follows.
These two methods are equivalent and one finds the following non-trivial relation
\begin{align}
	(R^{-1}\dmu R)_{ij} = \delta_{ij} 2Y_i^2 -2C_{ji},
\end{align}
where $C_{ij}$ is defined in \eq{defCij}.

\subsection{Standard versus minimal}
\label{StandardvsMinimal}

To go from the standard scheme in \sec{sectstandard} to the minimal scheme in \sec{sectminimal}, all one has to do is to diagonalize the mass matrix with an orthogonal transformation.
At the bare level we have
\begin{align}
\label{defO}
O^{bT}_{ij}m_{jk}^{b2}O_{kl}^{b} =& \delta_{il}m_i^{b2}
\end{align}
and the bare fields are then related by $\phi_i^b = O_{ij}^b\Phi_j^b$.
The fermion field and mass are actually the same in both the schemes. 
However, note that the field is renormalized, $\Psi^b=Z_\Psi \Psi$, in the standard scheme.

Just as discussed in \sec{GeneralvsMinimal}, the schemes are related by field redefinitions and hence must be equivalent.
We can then relate the renormalized quantities like the bare ones, but with a ``renormalized'' rotation matrix $O_{ij}(\mu)$,
\begin{align}\label{defO}
\delta_{ij}m_i^2 =~& O_{ik}^Tm_{kl}^2 O_{lj}, &&& \Lambda_{ijkl} =& \lambda_{abcd} O_{ai} O_{bj} O_{ck} O_{dl}, \nn\\
Y_i =~& y_a O_{ai}. &&&&
\end{align}
The rotation matrix $O$ obeys a similar non-trivial $\mu$-dependence as the transformation in \sec{GeneralvsMinimal}.
Using \eq{defO} to get from the RGEs in the standard scheme in \eq{RGEstandard} to the RGEs in the minimal scheme in \eq{RGEminimal} leads to the relation
\begin{align}
	(O^T \dmu O)_{ij} = \Bigg\{\begin{array}{lr} 0, & \text{for } i=j,
	\\  2C_{ji} +2Y_iY_j, & \text{for } i\neq j.\end{array}
\end{align}

\section{Conclusions}

We have shown that even in the general case, having several scalar fields with the same quantum numbers, it is possible to choose a renormalization scheme such that there is no kinetic mixing of these fields. 
We have shown explicitly how this can be done at 1-loop order by defining three different schemes: the most general one allowing for all kinetic mixings; the minimal one where the only parameters and counterterms are the physically relevant ones; 
and the standard method where the fields and masses mix during renormalization which produces general counterterms. 
We showed that the three cases are related by scale dependent field redefinitions and also explicitly that the renormalization group equations are equivalent.

While the general scheme exhibits kinetic mixing terms in the Lagrangian, these are not present in the other two schemes.
The equivalent effect is instead encoded in other ways. 
In the minimal scheme the effect enters in the off-diagonal $C_{ij}$ terms that depend on the scalar masses.
In the standard scheme the fields mix under renormalization; which gives rise to non-diagonal anomalous dimensions.
Also the masses mix under renormalization and diagonalizing the mass-matrix after the RG evolution, to get the masses of the physical fields, redefines the fields in a way corresponding to the off-diagonal $C_{ij}$ terms in the minimal scheme. 

Another effect from the mixing of the fields under RG evolution in the standard scheme is that the basis changes during RG running. 
This is, for example, important for the case of 2HDMs with a broken \Zsym symmetry. 
To circumvent this, one either has to keep track of how the basis changes by also taking into account the anomalous dimensions of the fields and the evolution of the mass terms or alternatively only use basis-invariant quantities.

\section*{Acknowledgements}

This work is supported in part by the Swedish Research Council grants
contract numbers 2015-04089 and 2016-05996 and by
the European Research Council (ERC) under the European Union's Horizon 2020
research and innovation programme (grant agreement No 668679).

\bibliography{kineticmixing_paper}{}

\end{document}